# Exogenous Approach to Grid Cost Allocation in Peer-to-Peer Electricity Markets

T. Baroche, P. Pinson *Senior Member, IEEE*, R. Le Goff Latimier, H. Ben Ahmed

*Abstract*—The deployment of distributed energy resources, combined with a more proactive demand side, is inducing a new paradigm in power system operation and electricity markets. Within a consumer-centric market framework, peer-to-peer approaches have gained substantial interest. Peer-to-peer markets rely on multi-bilateral direct negotiation among all players to match supply and demand, and with product differentiation. These markets can yield a complete mapping of exchanges onto the grid, hence allowing to rethink our approach to sharing costs related to usage of common infrastructure and services. We propose here to attribute such costs in a number of alternative ways that reflects different views on usage of the grid and on cost allocation, i.e., uniformly and based on the electrical distance between players. Since attribution mechanisms are defined in an exogenous manner and made transparent they eventually affect the trades of the market participants and related grid usage. The interest of our approach is illustrated on a test case using the IEEE 39 bus test system, underlying the impact of attribution mechanisms on trades and grid usage.

*Index Terms*—Economic dispatch, Distributed optimization, Optimal power flow, Cost allocation, Product differentiation.

## I. INTRODUCTION

Distributed energy resources, jointly with ICT and energy system management for residential homes and buildings, are making us rethink our approach to power system operation. Especially, going down to lower levels of the network, new type of agents are appearing, namely prosumers, with the ability to produce and consume (and most likely store in a very near future). While substantial efforts are made to have power system operation evolve in view of that new context, electricity markets have not gone yet through the same process of accommodating this new context with its challenges and opportunities. Electricity markets are expected to go from producer-centric to consumer-centric [1], [2], while they will most likely include a peer-to-peer (P2P) and community-based component [3]. A P2P market relies on multi-bilateral direct trades among participants. Employing a P2P market framework could yield a number of advantages, for instance thanks to product differentiation and its consumer-centric nature, allowing for a wealth of new business models.

Product differentiation is to be understood here as the fact that market players can express preferences on the type and quality of the energy they will exchange. Such preferences could be for local energy generation, for energy with limited $CO_2$ emissions, etc. However, there may be discrepancies between market-clearing (and related dispatch) and feasible dispatch in view of grid-related and operational constraints. In parallel, while it appears normal to socialize grid-related costs in the current wholesale-retail market structure, a future with peer-to-peer exchange and preferences may allow to rethink the way we attribute such costs. Our objective here is hence to describe a consumer-centric market allowing to allocate grid-related costs in an exogenous manner. The various attribution mechanisms are to impact trades and subsequent network usage.

The first approach to coordinated multi-lateral electricity trades was already proposed nearly 30 years ago [4]. The original aim was to allow for the separation of economics and reliability of system operation, as is the case for current European pool-based electricity markets. The proposal involved an iterative process with all players proposing their trades first, followed by the system operator to decide whether the trades respect operational constraints, or not. This proposal was recently revited in [5], also analysing game-theoretical properties of the solutions obtained. In both cases, the authors pointed at the fact that charges for network usage were not considered. A second approach may consist in relying on optimal power flow (OPF) models, allowing to consider network constraints in an endogenous manner (see e.g. [6]). While those are traditionally solved in a centralized fashion, many decomposition techniques were proposed to also solve them in a distributed fashion [7]–[11]. More recently, works like that of [12] proposed to account for network limits in the presence of distributed renewable resources and using decentralized consensus on a blockchain. Even though those operational problems are increasingly considered in decentralized manner, these do not comprise a market construct while they do not account for how grid-related usage costs would be attributed.

We propose here an alternative to account for network constraints and attribution of network usage costs in a consumer-centric framework. Contrary to the previous approaches, networks constraints are not enforced directly. It is rather chosen to use incentives, in the form of network usage charges to influence the outcome of P2P markets. The resulting P2P market formulation comprises a simple tool for system and market operators to limit potential detrimental effects that might be induced by P2P markets on power networks.

The manuscript is structured as following. Firstly, a novel P2P market formulation is proposed in Section II. It is eventually solved in a distributed manner, though enabling to account for centrally designed market policies, e.g., in the form of

T. Baroche (corresponding author) and R. Le Goff Latimier , and H. Ben Ahmed are with the SATIE Laboratory located at Ecole Normale Superieure de Rennes, France, e-mail: thomas.baroche@ens-rennes.fr, roman.legoff-latimier@ens-rennes.fr, hamid.benahmed@ens-rennes.fr.

P. Pinson is with the Center for Electric Power and Energy, Danmarks Tekniske Universitet, Denmark, e-mail: ppin@dtu.dk.

The authors are partly supported by the Danish Innovation Fund and the ForskEL programme through the projects '5s' - Future Electricity Markets (12-132636/DSF), CITIES (DSF-1305-00027B) and The Energy Collective (grant no. 2016-1-12530).

network usage charges. Subsequently, Section III introduces three approaches to the design of network usage charges to enter the P2P market negotitation mechanisms through a generalized formulation allowing for product differentiation. Simulation results are presented and discussed in Section IV using a test case relying on the IEEE 39-bus system. Section V gathers our conclusions and perspectives for further work.

## II. P2P Market Design and Corresponding Optimization Problem

A P2P market is based on a community of agents with flexible consumption or production. Prosumers will not be explicitly considered. Agents will either be net producers (and gathered in a set $\Omega^p$) or net consumers (gathered in $\Omega^c$). Emphasis is eventually placed on a deterministic clearing mechanism for a single market time unit. It may readily be extended to multiple time units with temporally binding constraints, while uncertainty could also be considered in a scenario-based stochastic optimization framework. First, a novel distributed P2P market structure is described, then translated into a consensus + innovation algorithm inspired by, e.g., [9], [13].

### A. Problem Formulation

We first start by formulating the complete problem with network constraints, which we will simplify in the following. Let us assume that each agent has a quadratic cost function, as is common in the literature, e.g., [7], [9] for economic dispatch and optimal power flow problems. For agent $n$, this cost function reads

$$f_n(P_n) = \frac{1}{2} a_n (P_n)^2 + b_n P_n + c_n \quad (1)$$

where $a_n, b_n, c_n \geqslant 0$ are cost function's parameters and $P_n$ is the net power production or consumption of agent $n$, being positive for producers and negative for consumers.

In a classical economic dispatch problem in a market environment the goal is to maximize social welfare as the differential between consumer and producer loss functions, for a total community $\Omega$ of $N$ market players, yielding our objective function (3a). Net powers are optimized within a flexibility range defined by a lower $\underline{P_n}$ and an upper $\overline{P_n}$ bound, as expressed in (3e). Considering multi-bilateral trades calls for a split of net powers, in the manner of [13], into a set of multiple bilateral trades $P_{nm}$. Every possible bilateral power trades within the community can be condensed in a matrix $\mathbf{P}$ such that

$$\mathbf{P} = \begin{pmatrix} P_{11} & \cdots & P_{1N} \\ \vdots & \ddots & \vdots \\ P_{N1} & \cdots & P_{NN} \end{pmatrix} \quad (2)$$

where $P_{nm}$ is equal to zero if agent $m$ is not in agent $n$'s trading partnership set $\omega_n$. Net power are then obtained by $P_n = \sum_{m \in \omega_n} P_{nm}$. $\mathbf{P}$ is skew-symmetric (3b) to insure power balance of each trade, so $P_{nn} = 0$. Its allows to potentially differentiate prices of individual trades.

Accounting for the electrical network requires additional constraints. First, reactive power injections $\mathbf{Q} = [Q_1, \cdots, Q_N]$ are to be within a certain range (3f). Reactive power injections help regulating amplitudes of voltages $V_i$ of each node $i \in \mathcal{N}$ of the power system within regulatory bounds (3g). Power balance at each node must be respected (3i). Finally, lines have maximal current capacities (3h). The global endogenous problem is as follows

$$\min_{(\mathbf{P},\mathbf{Q})} \quad \sum_{n \in \Omega} f_n(P_n) \quad (3a)$$

$$\text{s.t.} \quad P_{nm} = -P_{mn} \quad (3b)$$

$$P_{nm} \geqslant 0 \quad \text{if } n \in \Omega^p \quad (3c)$$

$$P_{nm} \leqslant 0 \quad \text{if } n \in \Omega^c \quad (3d)$$

$$\underline{P_n} \leqslant P_n \leqslant \overline{P_n} \quad (3e)$$

$$\underline{Q_n} \leqslant Q_n \leqslant \overline{Q_n} \quad (3f)$$

$$\underline{V_i} \leqslant |V_i| \leqslant \overline{V_i} \quad i \in \mathcal{N} \quad (3g)$$

$$|I_{il}| \leqslant \overline{I_{il}} \quad (i,l) \in \mathcal{L} \quad (3h)$$

$$\sum_{n \in \mathcal{N}_i} P_n + jQ_n = V_i \sum_{(i,l) \in \mathcal{L}} I_{il}^* \quad i \in \mathcal{N}, \quad (3i)$$

where $I_{il} = Y_{ii} V_i + Y_{il} V_l$ is the current on the line from node $i$ to node $l$ and $Y_{il}$ its admittance. A direct resolution of this problem would constitute an endogenous approach of grid cost allocation. However the feasibility space of this problem is strongly non-convex, and not necessarily a single connected space. In addition, the coupling between market and grid constraints imply an intense involvement of the system operator at each step of the solving algorithm.

A first simplification to dissociate the market from grid considerations is to include network constraints in a regularization function $h$, equal to 0 if they are respected and $+\infty$ if they are violated. It can be noted that in this case $h$ depends on real and reactive power injections. Reactive injections will be assumed sufficient to compensate voltage drops. The focus is then put on the real power market. In this case $h$ only depends on real power and can be noted $h(\mathbf{P})$. The problem becomes

$$\min_{\mathbf{P}} \quad \sum_{n \in \Omega} f_n(P_n) + h(\mathbf{P}) \quad (4a)$$

$$\text{s.t.} \quad P_{nm} = -P_{mn} \quad (4b)$$

$$\underline{P_n} \leqslant P_n \leqslant \overline{P_n} \quad (4c)$$

$$P_{nm} \geqslant 0 \quad \text{if } n \in \Omega^p \quad (4d)$$

$$P_{nm} \leqslant 0 \quad \text{if } n \in \Omega^c. \quad (4e)$$

This paper aims at replacing the regularization function to result in an exogenous approach used as a compensation mechanism for operation costs such as maintenance, power losses, power injection compensations, etc. Allocating these expenses will allow the system operator to reach cost recovery. Production differentiation formulation introduced in [13] seems a good candidate for this purpose. The regularization function evaluating network constraints can be replaced by

$$h(\mathbf{P}) = \sum_{n \in \Omega, m \in \omega_n} \gamma_{nm} P_{nm} + \tilde{h}(\mathbf{P}) \quad (5)$$

where parameters $\gamma_{nm}$ are product differentiation prices and $\tilde{h}(\mathbf{P})$ is an additional policy. Parameter $\gamma_{nm}$ is used to impact the trade of agent $n$ with agent $m$ for its use of the electrical network. This cost allocation is separable among participants, and will be integrated in their objective function as it will be further discussed in section III. However, $\tilde{h}(\mathbf{P})$ might not be separable if it represent a global objective of the community.

In consequence, the resolution can be distributed but not decentralized. The presence of the system operator in the loop is still mandatory in such a formulation. When distributed, at each iteration the system operator possesses a copy, noted $\mathbf{Z}$, of all power trades. This global variable can contribute to power trades consensus. For this, reciprocity constraint (4b) is replaced by power consensus constraint (6b). Power trades' reciprocity is imposed by the system operator on the global variable with $\mathbf{Z} = (\mathbf{P} - \mathbf{P}^{\mathrm{T}})/2$, where $.^{\mathrm{T}}$ is the transpose function. The final problem is

$$\min_{\mathbf{P},\mathbf{Z}} \sum_{n \in \Omega} \left[ f_n(P_n) + \sum_{m \in \omega_n} \gamma_{nm} P_{nm} \right] + \tilde{h}(\mathbf{Z}) \quad (6a)$$

$$\text{s.t.} \quad P_{nm} = Z_{nm} \quad (6b)$$

$$\underline{P_n} \leqslant P_n \leqslant \overline{P_n} \quad (6c)$$

$$P_{nm} \geqslant 0 \quad \text{if } n \in \Omega^p \quad (6d)$$

$$P_{nm} \leqslant 0 \quad \text{if } n \in \Omega^c. \quad (6e)$$

For the rest of this paper no other global policies will be considered, so $\tilde{h}(\mathbf{Z}) = 0$.

### B. Distributed P2P Market: Consensus + Innovations

As presented in [7], [14] several distributed or decentralized algorithms are adapted to solve problem (6). As in [9], the consensus and innovations algorithm used is based on obtaining a distributed iterative solution of the Karush-Kuhn-Tucker (KKT) first order optimality conditions of the problem. The consensus and innovations algorithm presents the advantage of having light updates requiring only basic operators, which makes it adapted for small computation units – such as for households – and rather straight forward for an implementation.

The Lagrangian of (6) is

$$L(\mathbf{P}, \mathbf{Z}) = \sum_{n \in \Omega} \begin{bmatrix} f_n(P_n) + \sum_{m \in \omega_n} \gamma_{nm} P_{nm} \\ - \sum_{m \in \omega_n} y_{nm}(P_{nm} - Z_{nm}) \\ - \overline{\mu_n} \left( \overline{P_n} - P_n \right) \\ - \underline{\mu_n} \left( P_n - \underline{P_n} \right) \\ - \sum_{m \in \omega_n} (v^p_{nm} - v^c_{nm}) P_{nm} \end{bmatrix} \quad (7)$$

where $y_{nm}$, $\overline{\mu_n}$, $\underline{\mu_n}$, $v^p_{nm}$, $v^c_{nm}$ are respectively the Lagrangian multipliers of power consensus, upper and lower bounds and sign constraints (6b)-(6e). When applied to (6) the consensus and innovations iterative process aims at reaching a consensus on Lagrangian multipliers $y_{nm}$ of the power consensus constraint (6b). Multipliers $y_{nm}$ are interpreted as the prices of the commodity and are used as medium for negotiation. For each partnership of $\omega_n$, agent $n$ will estimate the optimal Lagrange multiplier for this trade by following the iterative process

$$y_{nm}^{k+1} = y_{nm}^k - \beta^k (y_{nm}^k - y_{mn}^k) - \alpha^k (P_{nm}^k - Z_{nm}^k) \quad (8)$$

where $\alpha^k$ and $\beta^k$ are weighing coefficients meeting the same conditions than in [9]. These estimates are sought to meet asymptotically through consensus. The power consensus constraint (6b) for this trade is sought to be verified asymptotically too through the innovations part of the price update.

The corresponding first order KKT stationarity condition on power trade $P_{nm}$ is

$$f'_n - y_{nm} + \gamma_{nm} + \overline{\mu_n} - \underline{\mu_n} - v^p_{nm} + v^c_{nm} = 0, \quad (9)$$

where $f'_n$ is the gradient of agent $n$'s cost function.

To converge towards this condition it is possible to consider two steps. The first step correspond to solving the problem without sign constraints. And the second step enforces them. At first agent $n$ updates its lower $\underline{\mu_n}^{k+1}$ and upper $\overline{\mu_n}^{k+1}$ power boundary Lagrangian multipliers such as follows

$$\overline{\mu_n}^{k+1} = \left| \overline{\mu_n}^k + \rho^k (Z_n^k - \overline{P_n}) \right|^+ \quad (10a)$$

$$\underline{\mu_n}^{k+1} = \left| \underline{\mu_n}^k + \rho^k (\underline{P_n} - Z_n^k) \right|^+ \quad (10b)$$

where $\rho^k$ is a tuning parameter and $|\,.\,|^+$ the positive part operator. Agents update their power consumption/production according to the new prices estimates $y_{nm}^{k+1}$ and new power boundaries Lagrangian multipliers. These updates are made under the assumption that cost functions $f_n$ have a bijective gradient. The asumption is valid as long as quadratic coefficients $a_n$ are different from zero. In the case of a producer power trades estimates are updated as follows

$$P_{nm}^{k+1} = \left| Z_{nm}^k + g_{nm}^k (f'^{-1}_n(y_{nm}^{k+1} - \gamma_{nm} - \overline{\mu_n}^{k+1} + \underline{\mu_n}^{k+1}) - Z_n^k) \right|^+ \quad (11)$$

where the positive part operator $|\,.\,|^+$ enforces power trade sign constraint. The negative part operator is used instead in the case of a consumer. The proposed gradient step factor $g_{nm}^k$ is as follows

$$g_{nm}^k = \frac{|Z_{nm}^k| + \tau k^{-\delta}}{\sum_{m \in \omega_n} (|Z_{nm}^k| + \tau k^{-\delta})} \quad (12)$$

with $\tau$ and $\delta$ are tuning parameters.

After collecting all power trade set points, the central agent deals with the global variable to facilitate the consensus. Global variable's update is as follows

$$Z_{nm}^{k+1} = \frac{P_{nm}^{k+1} - P_{mn}^{k+1}}{2}, \quad \forall (n,m) \in \Omega \times \omega_n. \quad (13)$$

Note that in the context of privacy it would be interesting to use a secured mechanism, as in [12], for the global variable update as well as prices updates.

### III. EXOGENOUS GRID COST ALLOCATION

When the goal is to obtain a P2P market with allocation of grid costs it is possible to use product differentiation as presented in (5). It may be noted that in such case product differentiation prices $\gamma_{nm}$ would be provided by the system operator rather than chosen by agents. A cost allocation policy refers to the way product differentiation prices are designed to allocate costs.

When an incident occurs on the network security dispositions are applied automatically. However, the market as defined initially do not intrinsically take this into consideration. It is possible to vary partnership sets $\omega_n$ but this would require special signals from the system operator to alert market participants. Product differentiation prices used as transmission charges might offer an indirect solution to handle deteriorated situations. Cost allocation policy could incite agents to shift from their usual partners to others not affected by the malfunctions without changing their routine. The corresponding cost allocation policy would then need to enable market islanding. This second requirement corresponds to a security market procedure with the least grid stress while waiting for repairing. This deteriorate mode could also be used in case of congestion, for example by pushing consumers to lessen their consumption.

After expressing what amount of money is collected by the system operator, three cost allocation policies are proposed.

### A. Total Fees

The money paid (resp. received) by agent $n$ for buying electricity from (resp. selling to) agent $m$ is given by the perceived price $\hat{y}_{nm} = y_{nm} - \gamma_{nm}$. Product differentiation prices are costly exogenous parameters. Hence, when agent $n$ is a consumer $\gamma_{nm}$ is negative, so the price paid $\hat{y}_{nm}$ is higher than the electricity price $y_{nm}$. For a producer the price received is lower than the electricity price, since its parameters $\gamma_{nm}$ are positive. The total money paid or received by agent $n$ is expressed by

$$Y_n = \sum_{m \in \omega_n} \hat{y}_{nm} P_{nm}. \quad (14)$$

The part reserved for the system operator is

$$\Gamma_n = \sum_{m \in \omega_n} \gamma_{nm} P_{nm}. \quad (15)$$

From the system operator's point of view, the total amount of money collected through product differentiation prices is

$$\Gamma_{\text{SO}} = \sum_{n \in \Omega} \sum_{m \in \omega_n} \gamma_{nm} P_{nm}. \quad (16)$$

As mentioned previously, this money can be used to cover operation expenses, such as maintenance, power losses, power injection compensations, etc.

### B. Unique Cost Allocation Policy

The simplest cost allocation policy is to share costs equally between members of the community. A unique network fee $u^{\text{uniq}}$ serves as an adjusting parameter for the system operator. In this framework no discrimination is made between participants. Because of the universality of this policy, agents in recurrently congested areas might not be spurred to behave in a responsible manner. In this case misbehavior of a few agents would penalize the rest of the community. If grid costs are equally shared between the buyer and the seller of a trade, product differentiation prices become

$$\gamma_{nm}^{\text{uniq}} = \pm \frac{u^{\text{uniq}}}{2}, \quad \forall (n,m) \in \Omega \times \omega_n, \quad (17)$$

where the sign of $\gamma_{nm}^{\text{uniq}}$ is such that $\gamma_{nm}^{\text{uniq}} P_{nm} > 0$ to be a costly criterion for agent $n$, so $> 0$ for producers and $< 0$ for consumers. The network fee $u^{\text{uniq}}$ is expressed in €/MW.

### C. Electrical Distance Cost Allocation Policy

To be more precise in how costs are allocated, it is possible to make product differentiation prices proportional to the electrical distance between agents. This cost allocation policy would incite agents to trade with their closest electrical partners. When costs are equally shared between trading partners and follow the same sign convention than previously, product differentiation prices become

$$\gamma_{nm}^{\text{dist}} = \pm \frac{u^{\text{dist}} \text{d}_{nm}}{2}, \quad \forall (n,m) \in \Omega \times \omega_n, \quad (18)$$

where $\text{d}_{nm}$ is the electrical distance between agent $n$ and agent $m$ and the distant network fee $u^{\text{dist}}$ an adjusting parameter. The network fee $u^{\text{dist}}$ is expressed in €/MW/distance unit.

The definition of an electrical distance is a crucial issue for this cost allocation policy. [15] recommends two electrical distances, developed to allow a better vulnerability assessment through a topological visualization of an electrical structure. It is possible to consider either

1) the *Thevenin Impedance Distance*, where each line is weighed by the norm of its Thevenin impedance after which a shortest path algorithm is performed to obtain the Thevenin electrical distance between two distant nodes, or
2) the *Power Transfer Distance*, where the absolute value of *Power Transfer Distribution Factors* induced by a single trade are summed.

The Thevenin impedance distance, only considering the shortest path, is more adapted to a radial network such as a distribution grid. On the other hand, the power transfer distance considering a dc power flow approximation of the entire network is better suited for a meshed network such as a transmission grid.

### D. Uniform Zonal Cost Allocation Policy

The unique cost allocation policy does not differentiate agents while the electrical distance one might individualize too much grid tariffs. Using a zonal cost allocation policy seems a good compromise between both. In this situation the electrical network would be divided in several zones assigned with a zonal network fee $u^{\text{zone}}$. These network fees serve as adjusting parameter. Each zone could be managed by a different system operator. Product differentiation prices are obtained by summing zonal network fees of trade's crossed zones. An easy way to define trade's crossed zones is to use the shortest path, as for the Thevenin electrical distance. The considered path is equivalent to the most probable or the easiest electrical path. Applying a very high price to a zone would incite agents not to trade with agents of this zone, and inside agents towards self-consumption. In this sense, the cost allocation policy allows to isolate a zone but only from the market point of view.

This mechanism is simplified by considering zonal network fees to be uniform. This way, the problem of how zonal network fees are design between zones is limited. For this uniform zonal cost allocation policy with costs shared equally between trading partners and with the same sign convention than previously, product differentiation prices become

$$\gamma_{nm}^{\text{zone}} = \pm \frac{u^{\text{zone}} \text{N}_{nm}^{\text{zone}}}{2}, \quad \forall (n,m) \in \Omega \times \omega_n, \quad (19)$$

where $\text{N}_{nm}^{\text{zone}}$ corresponds to the number of zone crossed by the trade. The network fee $u^{\text{zone}}$ is expressed in €/MW.

It may be noted that both electrical distance and zonal cost allocation policies depend on grid characteristics, which do not variate in time. The flexibility is given by the network fees, adapting product differentiation prices to grid status (e.g. between day and night).

## IV. Case-studies and Application Results

In this section the effects induced by the proposed cost allocation policies will be evaluated and compared to the results obtained in the case of a free market, with network fees equal to zero.

### A. Test Case

To evaluate market responses flexible agents need to be defined. In addition, to test feasibility of power commitments grid characteristics are needed. Many standards exist for OPF problems, such as IEEE 39 bus test system. However, these test systems do not take into account flexible loads, neither cost functions nor flexibility limits are defined for loads. Some test cases exist for P2P markets, such as in [9], but they do not take power system's characteristics in consideration.

There is a need for a new test case adapted to the study of P2P markets accounting for network constraints. The grid of the IEEE 39 buses test system is considered for which generators and loads are defines as flexible. Generators will keep the same power boundaries. A wide flexibility range is given to consumers, with bounds from 10% to 150% of IEEE test system's fixed loads. Quadratic cost function parameters, inspired from [9], are summarized in Table I, while line characteristics are those of the IEEE 39 buses test system. The final power network in Fig. 1 is divided in four administrative zones, arbitrary defined, and will be referred as the *New England* test case.

For example, when the consumer at node 16 buys power from the producer at node 39, cost allocation policies have different impact on product differentiation prices. Since this test case is based on a meshed network the power transfer distance is used. The power transfer electrical distance between node 16 and 39 is 7.3, without dimension. The Thevenin shortest path between the same nodes passes by nodes {16, 17, 18, 3, 2, 1, 39} which crosses 2 zones. This gives a ratio 7.3 between the electrical distance and uniform product differentiation prices. The ratio is of 2 when zonal product differentiation prices are compared with the unique ones.

### B. Free Market

The free market refers to the P2P market when not costs are allocated, network fees equal to zero. The free market in the *New England* test case leads to a consensus equilibrium for a free market electricity price of 58.1 €/MW. Independently

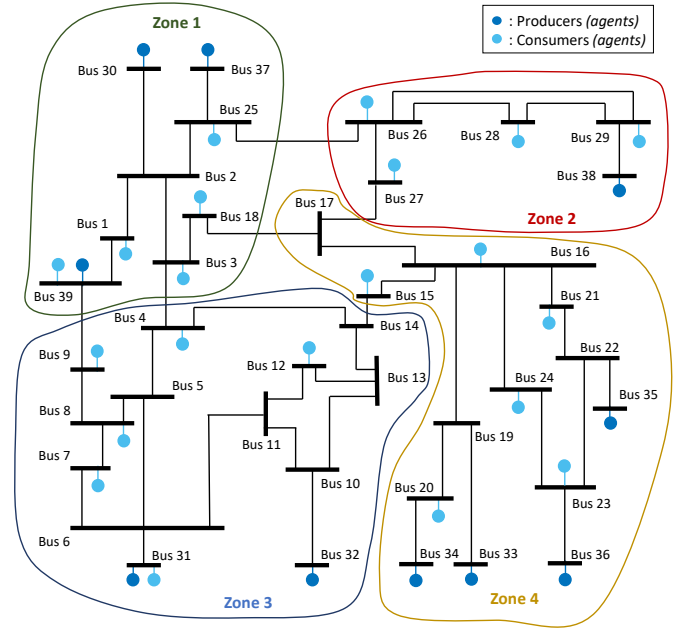

Fig. 1: New England test case for joint P2P market and OPF

TABLE I: Agents' characteristics of New England test case

| Agent | Bus | $a_n$ [€/MW$^2$] | $b_n$ [€/MW] | $\underline{P_n}$ [MW] | $\overline{P_n}$ [MW] |
|---|---|---|---|---|---|
| 1 | 1 | 0.067 | 64 | -146.4 | -9.76 |
| 2 | 3 | 0.047 | 79 | -483 | -32.2 |
| 3 | 4 | 0.047 | 71 | -750 | -50.0 |
| 4 | 7 | 0.053 | 62 | -350.7 | -23.38 |
| 5 | 8 | 0.082 | 65 | -783 | -52.2 |
| 6 | 9 | 0.052 | 83 | -9.8 | -0.65 |
| 7 | 12 | 0.087 | 63 | -12.8 | -0.853 |
| 8 | 15 | 0.057 | 81 | -480 | -32.0 |
| 9 | 16 | 0.050 | 73 | -493.5 | -32.9 |
| 10 | 18 | 0.052 | 69 | -237 | -15.8 |
| 11 | 20 | 0.071 | 62 | -1020 | -68.0 |
| 12 | 21 | 0.064 | 79 | -411 | -27.4 |
| 13 | 23 | 0.057 | 60 | -371.3 | -24.75 |
| 14 | 24 | 0.082 | 80 | -462.9 | -30.86 |
| 15 | 25 | 0.069 | 78 | -336 | -22.4 |
| 16 | 26 | 0.069 | 70 | -208.5 | -13.9 |
| 17 | 27 | 0.086 | 62 | -421.5 | -28.1 |
| 18 | 28 | 0.054 | 70 | -309 | -20.6 |
| 19 | 29 | 0.078 | 66 | -425.3 | -28.35 |
| 20 | 31 | 0.081 | 70 | -13.8 | -0.92 |
| 21 | 39 | 0.059 | 71 | -1656 | -110.4 |
| 22 | 30 | 0.089 | 18 | 0 | 1040 |
| 23 | 31 | 0.067 | 21 | 0 | 646 |
| 24 | 32 | 0.055 | 37 | 0 | 725 |
| 25 | 33 | 0.082 | 25 | 0 | 652 |
| 26 | 34 | 0.088 | 17 | 0 | 508 |
| 27 | 35 | 0.076 | 38 | 0 | 687 |
| 28 | 36 | 0.084 | 28 | 0 | 580 |
| 29 | 37 | 0.077 | 36 | 0 | 564 |
| 30 | 38 | 0.051 | 38 | 0 | 865 |
| 31 | 39 | 0.087 | 19 | 0 | 1100 |

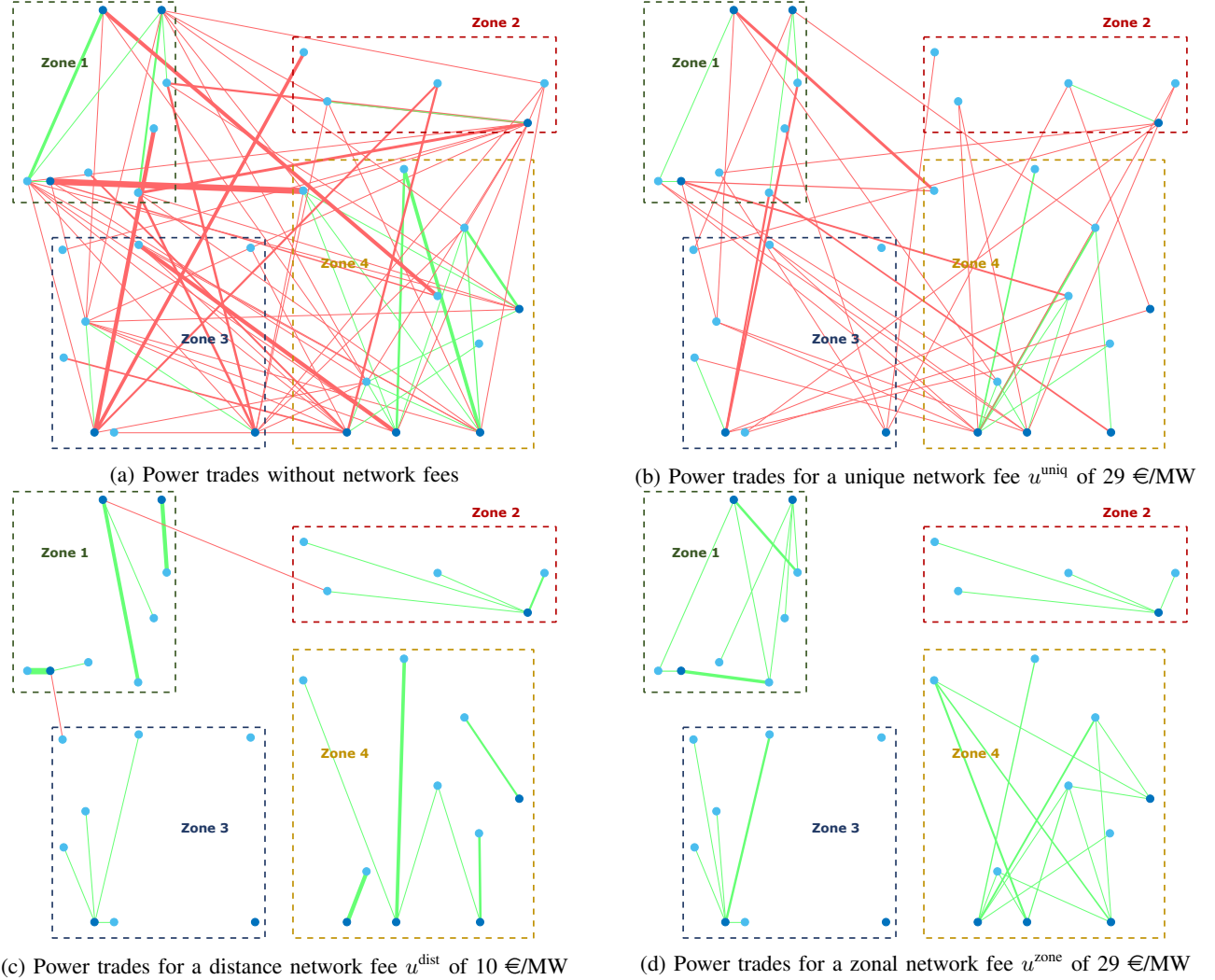

(a) Power trades without network fees

(b) Power trades for a unique network fee $u^{\text{uniq}}$ of 29 €/MW

(c) Power trades for a distance network fee $u^{\text{dist}}$ of 10 €/MW

(d) Power trades for a zonal network fee $u^{\text{zone}}$ of 29 €/MW

Fig. 2: Influence of network fee policies on market structure (red lines: inter-zone exchanges, green lines: intra-zone exchanges)

from the power network, it is important to study interactions between participants. This aspect will allow to verify if the conjectures made, when cost allocation policies have been defined, are correct. Looking at how trades are distributed between participants and the exchanges between zones seem also relevant. Moreover, the way the grid is used can be studied in a second step. This will help to point the presence of potential congestion.

To visualize resulting relations between agents it is possible to look into graph theory visualizations such as interaction diagrams. However, the interpretation of market interactions with such representation might be complex. A more intuitive visualization is to show the trades on power system's map Fig. 1. For clarity reasons it is recommended not to show lines and buses. Finally, to emphasize the difference between intra- and inter-zone exchanges, they will respectively be represented by green and red lines. To discriminate between main and small trades lines thickness will be proportional to the power traded. Only trades over $10^{-2}$ MW are represented in Fig. 2.

Fig. 2a shows power trade exchanges between market participants in the free market. Most trades are inter-zone exchanges. This implies that there is no correlation between the market and the power network restrictions, agents do not favor local trades. In consequence there is a high amount of power flowing between zones. The global absolute exchanges between zones is above 2 GW. A DC power flow analysis shows one congested line located between node 16 and node 19, used up to 130% of its capacity. Hence the risk of congestion does not originate from exchanges between zones.

### C. Impact on Peers' Interactions

It is then evident that no link exists between market operations and grid infrastructure. Understandably, feasibility of market outcomes is not guaranteed. Even though the commitments were applicable, the system operator would not be able to allocate operation costs without using product differentiation prices. The proposed cost allocation policies have a direct impact on relationships between community members. The increase of network fees might incite agents to trade less or differently.

As the unique cost allocation policy do not discriminate trades strong interactions between zones persist with a unique network fee $u^{\text{uniq}}$ of 50% of the free market electricity price, Fig. 2b. This policy does not encourage to adopt a cohesion between agents of the same area. However, the level of trades are much lower than the free market. The number of relevant trades also suffered from an important decrease. Some of the smallest consumers are only able to contract small power exchanges which make them appear isolated. The presented numerical result depends on the chosen characteristic for each agent. However the possibility to impact the grid power flow thank to a market fee is a general property.

With the electrical distance cost allocation policy the reduction of power trades level is reach much sooner. Already for a network fee of 17% of the free market electricity price the number of relevant trades has plummeted, Fig. 2c. This leaves more agents with small trades which reflects the fact that agents are at a lower level of flexibility than for the previous unique cost allocation policy, analysis deepen in next subsection. Power interactions has been modified to appear grouped by area. The market appears to be more coherence with grid infrastructures. It can be noted that only two inter-zone trades are still over $10^{-2}$ MW.

Finally, the zonal cost allocation policy can reach the same type of interactions with a fee of 50% of the free market electricity price, Fig. 2d. The pressure does not reach the same level of constraint on market's flexibility, which is translated by more relevant trades than in Fig. 2c. It can be noted that the zonal cost allocation policy avoids completely inter-zone exchanges.

To verify the conjecture that electrical distance and zonal cost allocation policies improve the coherence between market and network, it is possible to consider the variation of global absolute exchanges between zones depending on network fees, Fig. 3. The graph seems to mitigate this conjecture. As expected the amount of power traded between zones under a unique cost allocation policy seems to follow a linear decrease. It reaches a lower limit when market looses its flexibility. The lower bound being above zero indicates that the unique cost allocation policy does not push towards zonal self-consumption. This market configuration rather pushes agents to self-consume. With the zonal cost allocation policy the similar pattern is observed but with an ability to completely annihilate trades between zones. Trend's similarity actually originate from the uniform definition of zonal prices. If zones had different zonal prices, the zonal cost allocation policy was expected to be more discriminating while conserving its zonal islanding capability. Even though electrical distance cost allocation policy seems to be more efficient at first, the amount of power exchanged between zones is still important, coming from a high number of small trades.

### D. Effects on Power Network Usage

Market aggregation by zone does not reflect how the electrical grid is used. Cost allocation policies not only affects, but also provide an adjustment variable to the system operators. Network fees enable to tweak the expected outcomes of the P2P market in such way line capacities might not be violated and operation costs might be recovered by the system operator.

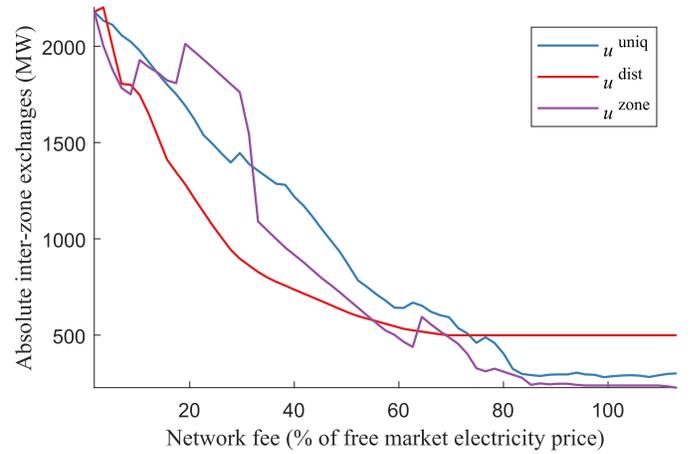

Fig. 3: New England test case for joint P2P market and OPF

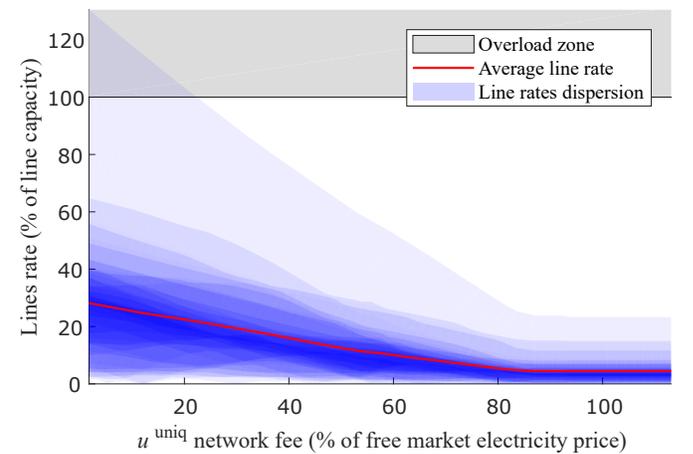

Fig. 4: Line rates of market outputs under unique network fees

Looking at the power flows induced by power commitments is more relevant to evaluate the efficiency of cost allocation policies. The market does not follow any physical limitations unlike the electrical grid, thus the difference between a feasible and none feasible market equilibrium lies in power network's feasibility set. In Fig. 4 DC power flows are represented for different unique network fees by steps of 1 €/MW. For this market community in average lines are used way below their capacity. However, one line is overloaded. As expected when the network fee increases, lines usage decreases. In this case, the most solicited line usage falls below its capacity limit. The absolute disparity of line rates drops till market's flexibility is lost.

To compare the three cost allocation policies only the average (continuous) and the maximum (dashed) line rates are plotted in Fig. 5 (upper part). It can easily be seen that the electrical distance cost allocation policy behaves differently from the other two. As an indication, a zone corresponding to the possible range of the average line rate is shown. The lower and upper bounds are respectively defined as the average line rate for the global minimum and maximum consumption.

Both maximum and average line rates decrease linearly with the unique and zonal network fees. Their behavior is linear because they are defined uniformly or uniformly over zones.

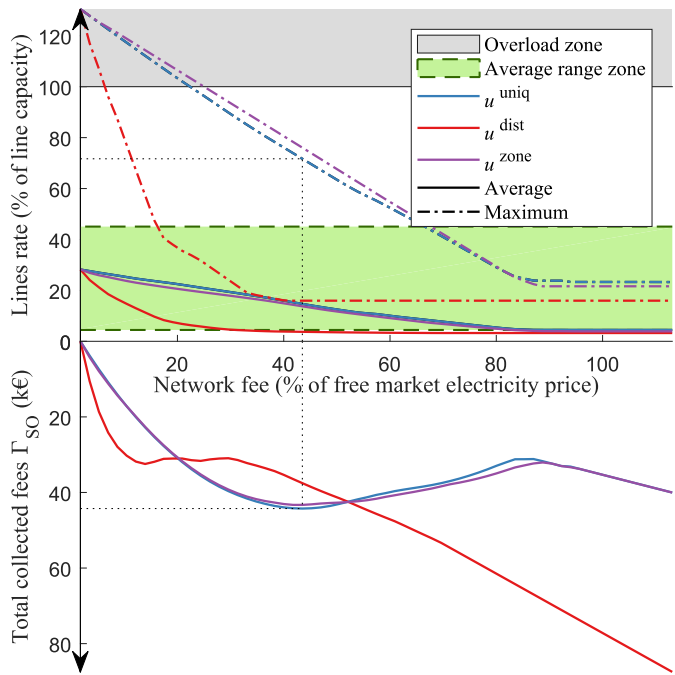

Fig. 5: Consequences of network fee policies over line rates (upper part) and total money collected for cost recovery (lower part)

As noted in subsection IV-B the most stressed line is inside a zone. In consequence, the zonal cost allocation policy is not able to remedy the congestion in a better way than the unique cost allocation. The electrical distance cost allocation policy generates a greater impact on the market which is translated in a faster decrease not only on the average but also on the maximum line rate. The electrical distance cost allocation policy allows to obtain feasibility of market commitments with the lowest over-cost for market participants. However, the money collected by the system operator might not be sufficient in this case to reach cost recovery.

The lower part of Fig. 5 represents the total amount of money collected by the system operator. From these curves it is clearly possible to identify when the market is too constrained. In this case, power trades stay the same while network fee criterion continue to increase, which leads to a linear increase. It can be observed that given market flexibility a local maximum exists, dashed lines for the unique cost allocation policy. The system operator can then deduce if operation costs can be recovered for a given cost allocation policy.

Finally, Fig. 5 offers a graphic tool for the system operator. It can choose which network fee to apply, for a given cost allocation policy, by following the proposed guide. For a given maximum acceptable line rate the corresponding network fee and the amount of money collected can be deduced. Or, for a given amount of money to collect to recover operation costs the corresponding network fee and maximum line rate can be deduced.

## V. Conclusion

Peer to peer markets are considered as a likely evolution of the power systems driven by distributed energy resources and ICT. In this paper a peer to peer electricity market including product differentiation has been considered. Product differentiation terms have been used as incentives to account for grid operation costs on which the market relies on. This mechanism incites market participants to respect power system's limits rather than enforcing them. Tested on three incentive frameworks, with a test case based on the IEEE 39 bus test system, it has been shown the ability of this mechanism to limit market outcomes' stress on the physical grid. Product differentiation terms also allow the system operator to collect money from market participants for their use of the grid.

This exogenous approach is a candidate for an applicable future step peer to peer market and is extendable to multiple time step considerations. The next step towards this is to compare the actual efficiency of this exogenous approach to an endogenous one, which guarantees to be applicable on the grid. If proven efficient, an online learning mechanism could be develop to obtain incentive parameters, provided a priori, in a robust way.